\documentclass[aps,pre,preprint,floatfix,groupedaddress]{revtex4-1}
\usepackage{graphicx}   			             		
\usepackage[parfill]{parskip}	
\usepackage{amssymb}
\usepackage{color}
\usepackage{amsmath}
\usepackage{amsfonts}
\usepackage{lineno,hyperref}
\begin{document}
\title{The fully frustrated XY model revisited: A new universality class}
\author{A. B. Lima}
    \affiliation{Departamento de 'F\'{\i}sica Aplicada, Universidade Federal do Tri\^angulo Mineiro, Uberaba, Minas Gerais, Brazil}
\author{B. V. Costa}
    \affiliation{Laborat\'orio de Simulac\~ao, Departamento de F\'{\i}sica, ICEx Universidade Federal de Minas Gerais, 31720-901 Belo Horizonte, Minas Gerais, Brazil}
\begin{abstract}
    The two-dimensional ($2d$) fully frustrated Planar Rotator model on a square lattice has been the subject of a long controversy due to the simultaneous $Z_2$ and $O(2)$ symmetry existing in the model. The $O(2)$ symmetry being responsible for the Berezinskii - Kosterlitz - Thouless transition ($BKT$) while the $Z_2$ drives an Ising-like transition. There are arguments supporting two possible scenarios, one advocating that the loss of $Ising$ and $BKT$ order take place at the same temperature $T_{t}$ and the other that the $Z_2$ transition occurs at a higher temperature than the $BKT$ one. In the first case an immediate consequence is that this model is in a new universality class. Most of the studies take hand of some order parameter like the stiffness, Binder's cumulant or magnetization to obtain the transition temperature. Considering that the transition temperatures are obtained, in general, as an average over the estimates taken about several of those quantities, it is difficult to decide if they are describing the same or slightly separate transitions. In this paper we describe an iterative method based on the knowledge of the complex zeros of the energy probability distribution to study the critical behavior of the system. The method is general with advantages over most conventional techniques since it does not need to identify any order parameter \emph{a priori}. The critical temperature and exponents can be obtained with good precision. We apply the method to study the Fully Frustrated Planar Rotator ($PR$) and the  Anisotropic Heisenberg ($XY$) models in two dimensions. We show that both models are in a new universality class with $T_{PR}=0.45286(32)$ and $T_{XY}=0.36916(16)$ and the transition exponent $\nu=0.824(30)$ ($\frac{1}{\nu}=1.22(4)$).
\end{abstract}
\maketitle
\section{Introduction}
    It is well known since the work of Mermin and Wagner \cite{Mermin-Wagner} that in one and two dimensions, continuous symmetries cannot be spontaneously broken at finite temperature in systems with sufficiently short-range interactions . However, Berezinskii \cite{Berezinskii} and Kosterlitz and Thouless \cite{Kosterlitz-Thouless-1} have shown that a quasi-long-range order characterized by a change in the behavior of the two point correlation function at a temperature $T_{BKT}$ can still exist. Magnetic prototypes undergoing such transition are the Planar Rotor ($PR$) \cite{40years,Minhagen-1} or the Anisotropic Heisenberg Model \cite{Rocha-Mol-Costa-1} (Also known as $XY$ model) in two dimensions. The PR and the $XY$ models share the same hamiltonian formula $H=-J\sum_{<i,j>}{S^{x}_{i}S^{x}_{j}+S^{y}_{i}S^{y}_{j}}$. However, in the $PR$ model the spins are restricted to the circle $\vec{S} = |\vec{S}|\left(  \cos \theta ~ \hat{x} + \sin \theta ~ \hat{y} \right)$ while in the $XY$ model $\vec{S} = |\vec{S}|\left(  \sin \theta \cos \phi ~ \hat{x} + \sin \theta \sin \phi ~\hat{y} + \cos \phi ~ \hat{z} \right)$. They are in the same class-of-universality \cite{Figueiredo-Rocha-Costa-1}. The PR (and the $XY$) model is interesting in its own right, as well as being a model for 2d Josephson-junction arrays, liquid helium superfluidity films, rough transition and many others \cite{Minhagen-1}. The nature of this transition is completely different from the common discontinuous (First order) or continuous (Second order) phase transitions. The two point correlation function, $\mathcal{C}(r)$, at low temperature,$T \leq T_{BKT}$, has a power law decay, $\mathcal{C}(r) \propto r^{-\eta(T)}$,  while an exponential decay, $\mathcal{C}(r) \propto e^{- r/\xi(T)}$, takes over for $T > T_{BKT}$ \cite{Kosterlitz-Thouless-1,Rocha-Mol-Costa-1,Kenna-Irving}. A model displaying a $BKT$ transition has an entire line of critical points in the low temperature region. Beside, the correlation length is expected to diverge exponentially as long as $T_{BKT}$ is approached from above, i.e. $\xi \propto e^{b(T-T_{BKT})^{-\nu}} ~, ~ T \rightarrow T^{+}_{BKT}$. The renormalization group theory predicts, $\nu = 1/2$ and $\eta = 1/4$ at $T_{BKT}$. The correlation exponent is expected to be a function of temperature \cite{Kosterlitz-Thouless-1,Figueiredo-Rocha-Costa-1}. The corresponding free energy is a $C^\infty$ function, but not analytical in the $T \leq T_{BKT}$ region. Its phenomenology relies on the belief that it is driven by a vortex-antivortex unbinding mechanism \cite{Berezinskii,Kosterlitz-Thouless-1}. The two dimensional Fully Frustrated Planar Rotator ($FFPR$) model (Or the corresponding $FFXY$ model) is a continuum model with uniform frustration which was originally proposed as a version of magnetic systems possessing frustration without disorder \cite{Villain-1}, it is now known to describe a $2D$ 2d Josephson-junction array in a perpendicular magnetic field \cite{Teitel-Jayaprakash-1} with the strength of the magnetic field corresponding to one magnetic-flux quanta for every plaquette of the array to which corresponds a $Z_2$ symmetry besides the continuous spin symmetry. The phase transitions of this model on a square lattice have been the subject of a long controversy \cite{Teitel-Jayaprakash-1,Thijssen-Knops, Granato-Nightingale,Knops-Nienhuis-Knops-Blote,Nightingale-Granato-Kosterlitz,
    Nicolaides,Ramirez-Santiago-Jose,Granato-Kosterlitz-Nightingale-1,Grest,Lee,Lee-Lee,
    Olsson-1,Granato-Kosterlitz-Nightingale-2,Jose-Ramirez-Santiago,
    Hasenbusch-Pelisseto-Vicari,Lima-Costa}. As a matter of simplicity we will use $FF$ to refer to both model, FFPR and FFXY, except when the distinction is essential to the understanding. The hamiltonian describing the $FF$ model is customarily written as
\begin{equation}
    H_{FF} = -J \sum_{<i,j>} \cos \left( \theta_i - \theta_j + A_{ij}  \right)  ,
    \label{Hamiltonian-FF}
\end{equation}
\noindent
    with, $J > 0$. The frustration is determined by the gauge field $A_{ij}$. Full frustration corresponds to one-half quantum flux per plaquette, $\varphi$, which means that $\varphi = \pm \frac{1}{2} = \frac{1}{2\pi}\sum A_{ij}$, where the sum is around the plaquette. The ground state for this model on a square lattice has plaquettes with clockwise and counterclockwise rotation in a checkerboard pattern \cite{Villain-1}. This checkerboard pattern gives rise to the discrete $Z_2$ symmetry of the anti-ferromagnetic Ising model \cite{Onsager}. At low temperature therefore this model is expected to have both the topological quasi-long-range order of the $XY$ model, and the ordinary long-range order of the Ising model. As a consequence of frustration, the ground state of the $FF$ model presents an $O(2) \otimes Z_2$ degeneracy. While the $O(2)$ degeneracy is related to the global invariance of the Hamiltonian, the additional $Z_2$ degeneracy is related to the breaking of the lattice translational invariance. The simultaneous $Z_2$ and $O(2)$ symmetries lead to the interesting possibilities of two kinds of phase transitions: a $BKT$ and a Ising-like one.

    It has been observed that the $BKT$ and the Ising transitions occur at a very close, if not equal, temperatures. Because of that, the nature of the phase transition is rather inconclusive, in particular, there exists controversy as to whether the two transitions occur at the same or separately at two different temperatures. Monte Carlo transfer-matrix studies \cite{Thijssen-Knops} appear to point in the direction of critical exponents which differ significantly from those of a pure Ising model. These exponents are in agreement with those on the single transition line of the coupled $PR-Ising$ model \cite{Granato-Kosterlitz-Nightingale-1}, which suggests a single transition of a new universality class. This single-transition scenario has also been favored by Monte Carlo simulations of the $PR$ \cite{Nicolaides} and of the coupled $PR-Ising$ models \cite{Granato-Kosterlitz-Nightingale-1}.
    In contrast to this single-transition scenario, finite-size scaling analysis of Monte Carlo results has found double transitions in the Coulomb gas system of half-integer charges \cite{Grest,Lee}, which is believed to be in the same universality class as the $FF$ models. In particular, the higher temperature transition has been found to be of the different universality class from the pure Ising one, suggesting that the non-Ising exponents of the Ising-like order parameter may not be regarded as evidence for the single transition. High-precision Monte Carlo simulations of the $FFPR$ model \cite{Lee-Lee} has also led to two transitions at slightly different temperatures. Further, the chirality-lattice melting transition at the higher transition temperature was suggested to belong to a new universality class rather than to the Ising one. A recent argument that the previously obtained non-Ising exponents are artifacts of the invalid scaling assumption \cite{Olsson-1} has raised more controversy.  \\

    In all works cited above there is the necessity of defining an order parameter either to obtain the Ising or the $BKT$ transition. In the case, when the existence of a unique transition is certain, $T_{t}$, is calculated as the average between several estimates obtained using different quantities. In the present case, if two transitions are present and very close one of the other, the situation is subtler. Also, we have to consider that small deviations in determining $T_{t}$ are amplified in the determination of the critical exponents \cite{LivroLandau}. In this paper we analyse the transition in the $FF$ models under the perspective of a new technique, based on the partial knowledge of the zeros of the probability energy distribution \cite{Rocha-Mol-Costa-1}. The method has shown to locate the transition temperature with high precision , even in the case of two concurrent transitions as discussed in Ref. \cite{Costa-Mol-Rocha-1,Costa-Mol-Rocha-2}. Using the zeros of the probability energy distribution method we do not need to know an order parameter \emph {a priori}. The critical exponent $\nu$ is obtained independently, without the need to know the transition temperature in advance. Our study cover, the $PR$ and the $XY$ models. The results clearly show that there is only one transition temperature in both cases with $T_{PR}=0.45286(32)$ and $T_{XY}=0.36916(16)$. The transition exponent $\nu=0.824(30)$ ($\frac{1}{\nu}=1.220(40)$).
\section{Fisher Zeros}
    Fisher has shown how the partition function can be written as a polynomial in terms of the variable $z = e^{-\beta \epsilon}$, where $\beta=1/k_{B}T$ is the inverse of the temperature, $T$, $k_{B}$ is the Boltzmann constat, and $\epsilon$ is the energy difference between two consecutive energy states \cite{Fisher-1,Yang-Lee,Fisher-2}. For a finite system, all roots of the polynomial lie in the complex plane. The coefficients of the polynomial are real implying that their roots appear in conjugate pairs. If the system under consideration undergoes a phase transition at a temperature $T_{t}$, the corresponding zero, $z_{t}$, must be real in the thermodynamic limit. To make those statements clearer we recall that the partition function can be written as
\begin{equation}
    Z = \sum_{\textbf{E}} g\left( \textbf{E} \right ) e^{-\beta \textbf{E}} = e^{-\beta \varepsilon_0} \sum_{n=1}^{N} g_{\emph{n}} e^{-\beta \emph{n} \varepsilon} ,
    \label{Partition-Function}
\end{equation}
\noindent
    where it is assumed that the possible energies of the system, $\textbf{E}$, can be written as a discrete set $\left\{\textbf{E}_{n}=n\varepsilon \right\} ; n = 0, 1, 2,\cdots$ and $\varepsilon_0$ is some constant energy threshold. As pointed above, if the system undergoes a phase transition at $T_{t}$ the corresponding zero $z_{t}(L)$ moves toward the positive real axis as the system size grows. From now on we call it the dominant zero. In general if the system undergoes $M$ transitions we expect that the corresponding zeros $\{ z^\star(L) = a^\star(L) + \mathrm{i} b^\star(L) \}_M \in \{ z_j(L) \}_N$, will converge to the infinite volume limit $b^\star(L) \to 0$ as $L \to \infty$ while $\lim_{L \rightarrow \infty} a^\star(L) = a^\star(\infty)$.
\section{Energy Probability Distribution Zeros}
    If we multiply Eq.~\ref{Partition-Function} by $1 = e^{-\beta_0 \textbf{E}} e^{\beta_0 \textbf{E}}$ it is rewritten as
 \begin{equation}
    Z_{\beta_0} = \sum_{\textbf{E}} h_{\beta_0} \left( \textbf{E} \right) e^{-\textbf{E} \Delta \beta},
\end{equation}
\noindent
    where $h_{\beta_0} \left( \textbf{E} \right)=g \left( \textbf{E} \right)e^{-\beta_0 E}$ and $\Delta \beta = \beta - \beta_0$. Defining the variable $x = e^{-\varepsilon \Delta \beta }$ we obtain
\begin{equation}
    Z_{\beta_0} = e^{- \varepsilon_0 \Delta \beta  } \sum_{n} h_{\beta_0}({ \emph{n}}) x^n  ,
    \label{EPD_polynomial}
\end{equation}
\noindent
    where $h_{\beta_0}({\emph{n}})= h_{\beta_0} \left( \textbf{E}_n \right)$ is nothing but the non-normalized canonical energy probability distribution ($EPD$), hereafter referred to as the energy histogram at temperature $\beta_0$. There is a one to one correspondence between the Fisher zeros and the $EPD$ zeros. Constructing the histogram at the transition temperature, i.e., $\beta_0 = \beta_t$, the dominant zero will be at $x_t = 1$, i.e., $Z=0$ at the critical temperature ($\Delta \beta =0$) in the thermodynamic limit. For finite but large enough systems, however, a small imaginary part of $x_t$ is expected.  Indeed, we may expect that the dominant zero is the one with the smallest imaginary part on the real positive region regardless $\beta_0$. Once we locate the dominant zero its distance to the point $(1,0)$ gives $\Delta \beta$ and an estimate for $\beta_t$.
    For temperatures close enough to $\beta_t$ only states with non-vanishing probability to occur are pertinent to the phase transition. Thus, for $\beta_0 \approx \beta_c$ we can judiciously discard \emph{small} values of $h_{\beta_0}$. The dominant zero acts as an accumulation point such that even far from $\beta_c$ fair estimates can be obtained.
    With this in mind we can develop a criterion to filter the important region in the energy space were the most relevant zeros are located. The idea follows closely the well known \emph{Regula Falsi}  method for solving an equation in one unknown. The reasoning is as follows: We first build a normalized histogram $ h_{\beta_0^0}$ ($\textrm{Max}(h_{\beta_0^{0}})=1$) at an initial (False) guess $\beta_{0}^{0}$. Afterward, we construct the polynomial, Eq.~\ref{EPD_polynomial}, finding the corresponding zeros. By selecting the dominant zero, $x_t^0$, we can estimate the pseudo critical temperature, $\beta_t^0$. Regarding that $\beta_t(L)$ is the true pseudo-critical temperature for the system of size $L$, if the initial guess $\beta_0^0$ is far from $\beta_c(L)$ the estimative $\beta_t^0$ will not be satisfactory. Nevertheless, we can proceed iteratively making $\beta_0^1=\beta_t^0$, building a new histogram at this temperature and starting over. After a \emph{reasonable} number of iterations we may expect that $ \beta_{t}^{j} $ converges to the true $ \beta_t(L) $ and thus $x_t^j$ approaches the point $(1,0)$.  This corresponds to apply a sequence of transformations, $P$, such that $\beta^{n+1} = P \beta^{n}$. The transition temperature corresponds to the fixed point $\beta_t = P \beta_t$. The property $x_t^j \to (1,0)$ can be used as a consistency check in this iterative process. An algorithm following those ideas is:
\begin{enumerate}
    \item{Build a single histogram $ h_{\beta_0^j}$ at $\beta_0^j$.}
    \item{Find the zeros of the polynomial.}
    \item{Find the dominant zero, $ x_t^j$.}
	\subitem a) {If $ x_t^j$ is close enough to the point $(1,0)$, stop.}
	\subitem b) {Else, make $\beta_0^{j+1}=-\frac{\ln \left (\Re e\left \{x_t^j \right \} \right ) }{\varepsilon}+\beta_0^j$ and go back to 1.}
\end{enumerate}
\noindent
\begin{figure}
    \includegraphics[keepaspectratio,height=1.0\textwidth]{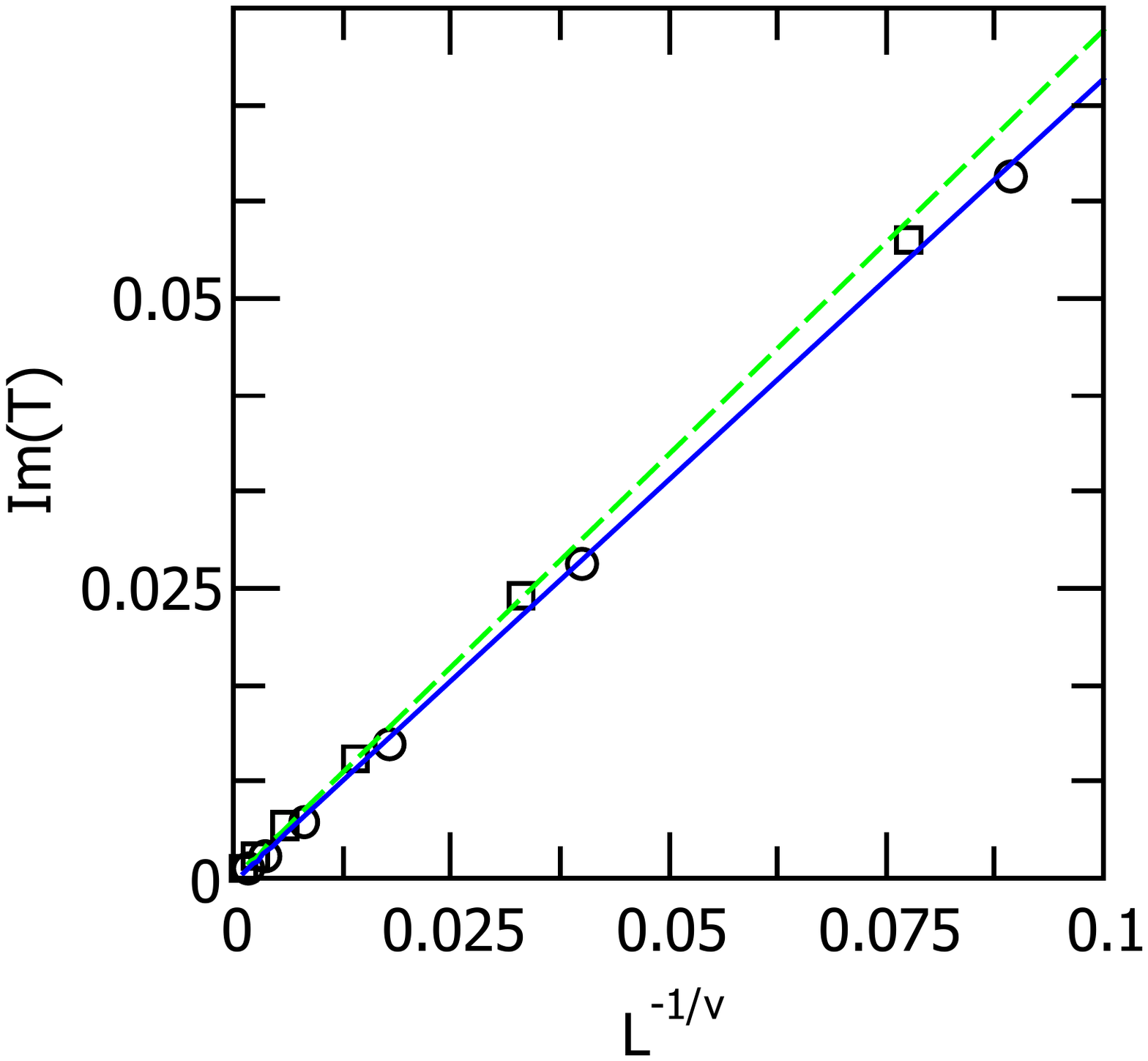}
    \caption{\label{Adjust_Im_T} Adjust of the real part of the dominant zeros for $L = 8, 16, 32, 64, 128$ and $256$. The circles (squares) and pluses (crosses) are for the $PR$ ($XY$) model with iterations coming from high and low temperatures respectively. The solid lines are adjusts using the minimum square method for $L > 16$.}
\end{figure}
\noindent
    In all our numerical results we observed that the choice of the starting temperature is irrelevant. To build the single histogram we follow the recipe given by Ferrenberg and Swendsen \cite{Ferrenberg-1,Ferrenberg-2}. It is noteworthy that if the system undergoes more than one transition the iterative procedure converges to the to the closer zero (Then the designation \emph{ dominant zero}) \cite{Costa-Mol-Rocha-1,Costa-Mol-Rocha-2}.
\section{Numerical details and results}
\begin{figure}
\includegraphics[keepaspectratio,height=1.0\textwidth]{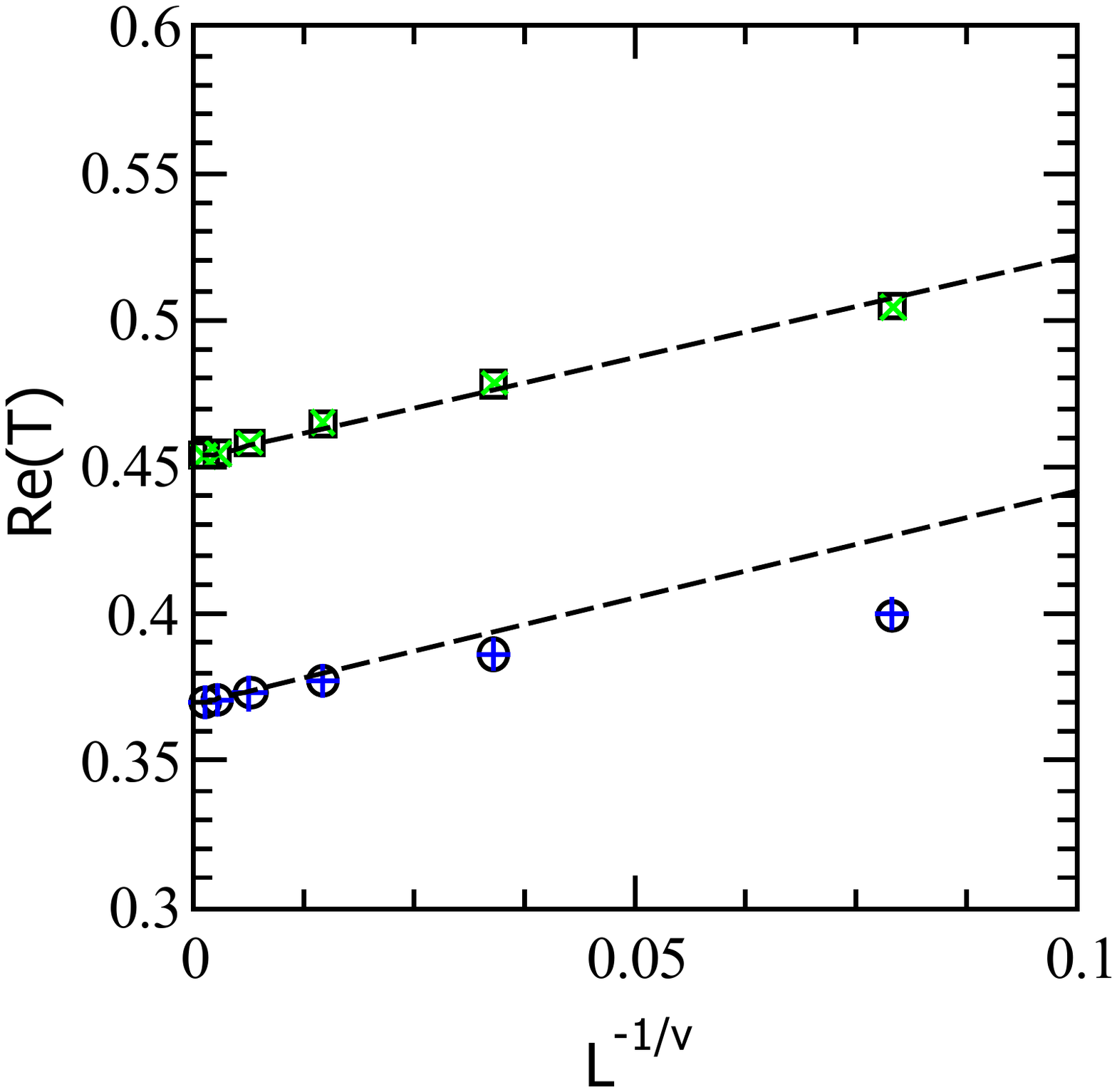}
    \caption{\label{Adjust_Re_T} Adjust of the imaginary part of the dominant zeros for $L = 8, 16, 32,64, 128$ and $256$. The data for $T^{low}(L)$ and $T^{high}(L)$ can hardly be distinguished due to the scale of the figure since differences are very small. The error bars are smaller than the symbols when not explicitly shown. The symbols are the same as those in Fig. \ref{Adjust_Im_T}}
\end{figure}
\noindent
    Let as suppose that the system has two transitions at temperatures $T_{-}$ and $T_{+}$ with $T_{-} < T_{+}$. As discussed in reference \cite{Costa-Mol-Rocha-1}, if we start the search at a temperature $T^{(0)} \mp \delta_{-(+)}$ the iterative procedure converges as $T^{n} \rightarrow T_{-(+)}$. Here $\delta_{-(+)}$ is a positive quantity. In general, the size of $\delta$ is not important, but as a matter of hastening the convergence we chose it closer to the transition temperature, when possible to guess it. A typical calculation is presented in Tab. \ref{Table-1}.
    In our simulations we have used a single spin Metropolis update discarding $100 \times L^{2}$ initial Monte Carlo steps (MCS) to reach equilibrium. Each histogram was built using $10^{9}$ configurations. Some care must be taken with the use of non-reliable pseudo-random number generator as discussed in Ref. \cite{Ferrenberg-Landau-Wong,Resende-Costa}. In the present case we have used the rannyu pseudo-number generator \cite{LivroLandau} as modified by Sokal  which has proven to be adequate here. To get the zeros we have used the package {\textbf{solve}} of the Mathematica\circledR program (Version 8). Our code was implemented using gfortran version 10.4.2 \cite{Gnu-Fortran}. Each point in our calculation is the result of the average over $4$ independent histograms. Error bars are smaller then the symbols in our figures when not explicitly shown.
\begin{table}[h]
    \caption{\label{Table-1}Typical table used to estimate the critical temperature for the $2d$ $FFPR$ model with the initial guess in the low temperature region. $\delta$ stands for the distance of the dominant zero to the $(1,0)$ point. Using $4$ independent histograms we obtain $T^{low}_{PR}{t}=0.45286(32)$ and $\frac{1}{\nu^{low}_{PR}}=1.236(41)$.}
\begin{center}
\begin{tabular}{cccccc}
  \hline
  L   & $\beta$     & $T$       & $\Re e(x)$  & $\Im m (x)$    	& $\delta$             \\
  \hline \hline
  8   & 0.25        & 4		    & 1.067586	  & 0.148106	    & 0.162798              \\
      & 0.298865	& 3.345998  & 1.054682	  & 0.114030	    & 0.126463              \\
      & 0.355416	& 2.813601  & 1.044631	  & 0.08160	        & 0.09301               \\
      & 0.420705	& 2.376962  & 1.029283	  & 0.05571	        & 0.06293               \\
      & 0.478851	& 2.088334  & 1.011730	  & 0.05493	        & 0.05617               \\
      & 0.507173	& 1.971714  & 0.998423	  & 0.05514	        & 0.05517               \\
      & 0.503146	& 1.987493  & 1.000807	  & 0.05515	        & 0.05515               \\
      & 0.505197    & 1.979424  &             &                 &                       \\
  \hline
  16  & 0.505197	& 1.979424	& 0.990391	& 0.022830	& 0.024769              \\
	  & 0.481702	& 2.075974	& 0.998392	& 0.024083	& 0.024137              \\
	  & 0.477996	& 2.092070	& 1.000137	& 0.024662	& 0.024663              \\
      & 0.478309	& 2.090697  &           &           &                       \\
  \hline
  32  & 0.478309	& 2.090738	& 0.995091	& 0.009763	& 0.010928              \\
	  & 0.467301	& 2.139950	& 0.998748	& 0.010406	& 0.010481              \\
	  & 0.464582	& 2.152475	& 0.999894	& 0.010088	& 0.010089              \\
	  & 0.464354	& 2.153530  &           &           &                       \\
  \hline
  64  &	0.464354    & 2.151926	& 0.997773	& 0.004530	& 0.005048              \\
	  & 0.459936	& 2.174218	& 0.999152	& 0.004475	& 0.004555              \\
	  & 0.458147	& 2.182706	& 1.000119	& 0.004585	& 0.004586              \\
	  & 0.458396	& 2.181520  &           &           &                       \\
  \hline
  128 & 0.458396	& 2.182929	& 0.998856	& 0.001964	& 0.002273              \\
	  & 0.455710	& 2.194377	& 0.999418	& 0.001859	& 0.001948              \\
	  & 0.454504	& 2.200200  &           &           &                       \\
  \hline
  256 & 0.454504	& 2.198237	& 0.998190	& 0.009713  & 9.879808              \\
	  & 0.454535 	& 2.200049	& 0.998560	& 0.009874	& 9.978543              \\
	  & 0.454238	& 2.201490	& 0.999085	& 0.007988	& 8.039690              \\
	  & 0.454049	& 2.202405  &           &           &                       \\
  \hline \hline
\end{tabular}
\end{center}
\end{table}
\noindent
\begin{table}[ht]
    \caption{\label{Table-2}This table presents the averaged values for the real and imaginary part of the pseudo-critical temperature for each lattice size $L$. The first and second entries for each $L$ are for the starting point above and below the expected transition temperature respectively as explained in the text. We remind the reader that $\Im m (T_{PR(XY)}) \rightarrow 0$ in the limit $L \rightarrow \infty$}
\begin{center}
\begin{tabular}{ccccc}
  \hline
  L   &$\Re e(T_{XY})$    & $\Im m (T_{XY})$   & $\Re e(T_{PR})$ & $\Im m (T_{PR})$   \\
  \hline \hline
 8    &   0.399050(39)	  &  0.0610(10)	       & 0.504810(17)	 &  0.05512(23)       \\
	  &   0.399906(10)	  &  0.0605(10)	       & 0.504300(13)	 &  0.05509(92)       \\
16	  &   0.386600(14)	  &  0.02719(32)	   & 0.478350(32)	 &  0.02434(15)       \\
	  &   0.386120(38)	  &  0.02715(2)        & 0.478500(33)	 &  0.02430(11)       \\
32	  &   0.377278(10)	  &  0.01111(31)       & 0.464780(19)	 &  0.01038(15)       \\
	  &   0.377580(30)	  &  0.01150(7)        & 0.465090(25)	 &  0.01031(10)       \\
64	  &   0.372900(29)	  &  0.00475(95)       & 0.458100(13)	 &  0.004525(33)      \\
	  &   0.372850(12)	  &  0.004827(15)      & 0.458170(13)	 &  0.000130(13)      \\
128	  &   0.370830(65)	  &  0.001898(24)      & 0.454910(18)	 &  0.000056(56)      \\
	  &   0.370698(14)	  &  0.001905(20)      & 0.454660(40)	 &  0.000058(58)      \\
256	  &   0.369916(16)	  &  0.000883(16)      & 0.454020(66)	 &  0.000042(42)      \\
	  &   0.369914(42)	  &  0.000835(55)      & 0.453820(11)	 &  0.000038(38)      \\
  \hline \hline
\end{tabular}
\end{center}
\end{table}
\noindent
\section{Final remarks}
    A decade ago Hasenbusch, Pelisseto and Vicari \cite{Hasenbusch-Pelisseto-Vicari} published a paper where they discussed in details the transition in the $FFPR$ model. In this paper they stated that
     `` Beside confirming the two-transition scenario, we have also observed an unexpected crossover behaviour that is universal to some extent. In the $FFXY$ model and in the $\phi^{4}$ and Is-XY ('Ising-XY') models, in a large parameter region, the finite-size behaviour at the chiral and spin transitions is model independent, apart from a length re-scaling. In particular, the universal approach to the Ising regime at the chiral transition is non-monotonic for most observable, and there is a wide region in which the finite-size behaviour is controlled by an effective exponent $\nu_{eff} \approx 0.8$. This occurs for $ L \leq \xi^{(c)}_{s}$, where $\xi^{(c)}_{s}$ is the spin correlation length at the chiral transition, which is usually large in these models; for example, $\xi^{(c)}_{s} = 118(1)$ in the square-lattice $FFPR$ model. This explains why many previous studies that considered smaller lattices always found $\nu \approx 0.8$.''
     Although the argument of Hasenbusch, Pelisseto and Vicari is sound, it should be interesting if we could confirm it using a different approach. In Tab. \ref{Table-2} we show the results of applying our method for the $FFPR$ and $FFXY$ models. It is noteworthy that in all entries the temperatures obtained for each size and model ($PR$ and $XY$) coincide within the error bars independent if they start above or below the estimated transition temperatures. This behavior is a clear indication that there is only one transition. The results are shown in Figs. \ref{Adjust_Re_T} and \ref{Adjust_Im_T}. If the opposite was to be true, we should obtain different temperatures in both cases for finite values of $L$ since the intermediate results do not depend on any finite size correction.  It is important to note that we know exactly the point $\Im m \{ T (L \rightarrow \infty ) \}=0$. This allows us to obtain the exponent $\nu$ from the imaginary part of $T$ without a previous knowledge of the transition temperature. The results we have obtained are fully consistent with a unique transition in a new universality class.
\section*{Acknowledgments}
    This work was partially supported by CNPq and Fapemig, Brazilian Agencies.
\end{document}